\documentclass[reprint,onecolumn,aps,pre,eqsecnum,longbibliography,showpacs]{revtex4-1}
\usepackage[latin9]{inputenc}
\usepackage{color}
\usepackage[english]{babel}
\usepackage{array}
\usepackage{textcomp}
\usepackage{amsmath}
\usepackage{amssymb}
\usepackage{graphicx}
\usepackage[unicode=true,pdfusetitle,
 bookmarks=true,bookmarksnumbered=true,bookmarksopen=true,bookmarksopenlevel=1,
 breaklinks=true,pdfborder={0 0 0},backref=false,colorlinks=true]
 {hyperref}

\makeatletter

\providecommand{\tabularnewline}{\\}

\@ifundefined{textcolor}{}
{%
 \definecolor{BLACK}{gray}{0}
 \definecolor{WHITE}{gray}{1}
 \definecolor{RED}{rgb}{1,0,0}
 \definecolor{GREEN}{rgb}{0,1,0}
 \definecolor{BLUE}{rgb}{0,0,1}
 \definecolor{CYAN}{cmyk}{1,0,0,0}
 \definecolor{MAGENTA}{cmyk}{0,1,0,0}
 \definecolor{YELLOW}{cmyk}{0,0,1,0}
 }

\usepackage{graphicx}

\DeclareGraphicsExtensions{.png,.pdf,.eps,.jpg}
\graphicspath{{.}}

\makeatother

\begin{document}

\title{The Non-Signalling theorem in generalizations of Bell\textquoteright{}s
theorem \vspace*{\bigskipamount}
}

\author{Jan \surname{Walleczek}\textsuperscript{1}}

\email[E-mail (corresponding author): ]{walleczek@phenoscience.com}

\author{Gerhard \surname{Grössing}\textsuperscript{2}\vspace*{\bigskipamount}
}

\email[E-mail: ]{ains@chello.at}

\affiliation{\textsuperscript{1}Phenoscience Laboratories, Chausseestrasse 8,
10115 Berlin, Germany }

\affiliation{\textsuperscript{2}Austrian Institute for Nonlinear Studies, Akademiehof\\
 Friedrichstr.~10, 1010 Vienna, Austria\vspace*{2cm}
}

\date{\today}
\begin{abstract}
\noindent Does ``epistemic non-signalling'' ensure the peaceful
coexistence of special relativity and quantum nonlocality? The possibility
of an affirmative answer is of great importance to deterministic approaches
to quantum mechanics given recent developments towards generalizations
of Bell\textquoteright{}s theorem. By generalizations of Bell\textquoteright{}s
theorem we here mean efforts that seek to demonstrate the impossibility
of any deterministic theories to obey the predictions of Bell\textquoteright{}s
theorem, including not only local hidden-variables theories (LHVTs)
but, critically, of nonlocal hidden-variables theories (NHVTs) also,
such as de Broglie-Bohm theory. Naturally, in light of the well-established
experimental findings from quantum physics, whether or not a deterministic
approach to quantum mechanics, including an emergent quantum mechanics,
is logically possible, depends on compatibility with the predictions
of Bell\textquoteright{}s theorem. With respect to deterministic NHVTs,
recent attempts to generalize Bell\textquoteright{}s theorem have
claimed the impossibility of any such approaches to quantum mechanics.
The present work offers arguments showing why such efforts towards
generalization may fall short of their stated goal. In particular,
we challenge the validity of the use of the non-signalling theorem
as a conclusive argument in favor of the existence of free randomness,
and therefore reject the use of the non-signalling theorem as an argument
against the logical possibility of deterministic approaches. We here
offer two distinct counter-arguments in support of the possibility
of deterministic NHVTs: one argument exposes the circularity of the
reasoning which is employed in recent claims, and a second argument
is based on the inconclusive metaphysical status of the non-signalling
theorem itself. We proceed by presenting an entirely informal treatment
of key physical and metaphysical assumptions, and of their interrelationship,
in attempts seeking to generalize Bell\textquoteright{}s theorem on
the basis of an ontic, foundational interpretation of the non-signalling
theorem. We here argue that the non-signalling theorem must instead
be viewed as an epistemic, operational theorem i.e.\ one that refers
exclusively to what epistemic agents can, or rather cannot, do. That
is, we emphasize that the non-signalling theorem is a theorem about
the operational inability of epistemic agents to signal information.
In other words, as a proper principle, the non-signalling theorem
may only be employed as an \textbf{epistemic}, \textbf{phenomenological},
or \textbf{operational} principle. Critically, our argument emphasizes
that the non-signalling principle must not be used as an ontic principle
about physical reality \textbf{as such}, i.e.\ as a theorem about
the nature of physical reality independently of epistemic agents e.g.\ human
observers. One major reason in favor of our conclusion is that any
definition of signalling or of non-signalling invariably requires
a reference to epistemic agents, and what these agents can actually
measure and report. Otherwise, the non-signalling theorem would equal
a general \textquotedblleft{}no-influence\textquotedblright{} theorem.
In conclusion, under the assumption that the non-signalling theorem
is epistemic (i.e.\ ``epistemic non-signalling''), the search for
deterministic approaches to quantum mechanics, including NHVTs and
an emergent quantum mechanics, continues to be a viable research program
towards disclosing the foundations of physical reality at its smallest
dimensions.
\end{abstract}
\maketitle

\section*{Introduction }

Historically, the first successful attempt in fashioning a deterministic
approach to quantum mechanics was the one now known as the de Broglie-Bohm
theory, and that theory introduced the notion of \textquotedblleft{}hidden
variables\textquotedblright{} (Bohm~\cite{Bohm.1952interpr1}). In
turn, Bohm\textquoteright{}s consideration that hidden variables might
possibly underlie quantum phenomena, in conjunction with the earlier
analysis by Einstein, Podolsky and Rosen~\cite{Einstein.1935can},
inspired John Bell to develop the theorem named after him (Bell~\cite{Bell.1964einstein}).
50 years on, it is widely accepted that Bell\textquoteright{}s groundbreaking
theorem has proven beyond doubt that no (deterministic) quantum theory
based on local hidden variables could account for the observations
of quantum physics. However, while Bell\textquoteright{}s theorem
ruled out as an explanation any local hidden-variables theory (LHVT),
Bohm\textquoteright{}s original proposal of a nonlocal hidden-variables
theory (NHVT) remained a viable possibility at least on logical grounds,
even after Bell (e.g.\ Bell~\cite{Bell.1987speakable}). 

In recent years, however, efforts have gained prominence that implied
that \textbf{all} types of deterministic approaches to quantum mechanics,
not only LHVTs, but also all forms of NHVTs such as de Broglie-Bohm
theory, must be in violation of Bell\textquoteright{}s theorem. In
the context of the present article, such efforts we will refer to
as efforts seeking to generalize Bell\textquoteright{}s theorem. Recent
publications with important implications for generalization, in the
here proposed sense, of Bell\textquoteright{}s theorem are, for example,
Colbeck and Renner~\cite{Colbeck.2011no,Colbeck.2012free} and Gallego
\textit{et al.}~\cite{Gallego.2013full}. (Their work builds on previous
work that has similar implications such as Barrett \textit{et al.}~\cite{Barrett.2005no};
Kofler \textit{et al.}~\cite{Kofler.2006experimenters}; Pironio
\textit{et al.}~\cite{Pironio.2010random}). Both Colbeck and Renner~\cite{Colbeck.2011no,Colbeck.2012free},
and Gallego \textit{et al.}~\cite{Gallego.2013full}, have argued
that even those deterministic quantum theories, which are unconstrained
by Bell\textquoteright{}s locality assumption (i.e.\ an NHVT such
as de Broglie-Bohm theory), must necessarily violate the non-signalling
theorem or principle. Gallego \textit{et al.}~\cite{Gallego.2013full}
also argued that their own work had important \textquotedblleft{}philosophical
and physics-foundational implications\textquotedblright{}. Because
of their interest in foundational issues as well and also because
this work is the most recent one we included in our analysis, we will
use Gallego \textit{et al.}~\cite{Gallego.2013full} as the key reference
in our exploration of basic physical and metaphysical assumptions
concerning the non-signalling theorem. While in no way do we question
the great technical achievements which are presented in the mentioned
articles, we do question the finality of assumptions expressed therein
concerning the possibility of deterministic approaches to quantum
mechanics. Importantly, these authors, like many researchers in quantum
physics in general, take for granted a foundational, ontic interpretation
of the non-signalling theorem. By contrast, the present work will
argue that for deterministic approaches to quantum mechanics an exclusively
\textbf{epistemic}, \textbf{phenomenological}, or \textbf{operational},
interpretation of the non-signalling theorem does in fact ensure compatibility
between special relativity and quantum nonlocality. Before proceeding
with our analysis of the non-signalling theorem, and of possible assumptions
underlying the theorem, we will start with a brief summary of quantum
phenomenology as apparent from EPR-type experiments.

\section{Quantum phenomenology of EPR-type experiments}

What is the phenomenon that is in need of explanation by quantum-theoretical
approaches whether by standard indeterminist or by unconventional
determinist approaches? We here focus on quantum phenomenology, such
as on nonlocal correlation phenomena, as apparent from EPR-type experiments
(e.g.\ Aspect \textit{et al.}~\cite{Aspect.1982experimental,Aspect.1982bell};
Weihs \textit{et al.}~\cite{Weihs.1998violation}; Tittel \textit{et
al.}~\cite{Tittel.1998violation}; Ursin \textit{et al.}~\cite{Ursin.2007free-space};
Giustina \textit{et al.}~\cite{Giustina.2013bell}). In the following
sections (1.1. to 1.3.) we present a brief overview of different metaphysical
interpretations of the same quantum-phenomenological observations.

\subsection{Nonlocal, instantaneous correlations signify the existence of instantaneous
influences}

Orthodox quantum theory makes no claim regarding the nature of physical
reality that might underlie the formation of nonlocal i.e.\ instantaneous
correlations. Everyone agrees that the term \textquoteleft{}correlation\textquoteright{}
merely represents a descriptive, epistemic term i.e.\ \textquoteleft{}correlation\textquoteright{}
defines a state of knowledge only. Therefore, the use of the term
neither implies underlying causal processes nor the existence of any
fundamental physical elements or ontic structures in general. However,
at the same time, we consider the fact to be non-controversial that
instantaneous correlations manifest \textbf{physically} at the level
of distant measuring devices. It goes without saying that lacking
the power to manifest physically, instantaneous correlations would
be beyond the capacity of epistemic agents to observe. Therefore,
the fact that instantaneous correlations can indeed be observed as
\textbf{physical effects} in concrete i.e.\ ontically-real systems,
we take as evidence that some kind of \textquotedblleft{}instantaneous\textquotedblright{}
or \textquotedblleft{}quasi-instantaneous\textquotedblright{} influence
must be at work as part of the formation, between space-like separated
quantum detectors, of nonlocal correlations. We here do not engage
the difficult question of whether the existence of \textquotedblleft{}nonlocal
correlations\textquotedblright{} is evidence of nonlocal or superluminal
\textbf{causation} (e.g.\ see Maudlin~\cite{Maudlin.2011quantum}).
In fact, it is of no consequence for our subsequent argument that
instantaneous influences \textendash{} at a minimum \textendash{}
represent near-instantaneous superluminal influences, whether or not
influences are assumed to be \textquotedblleft{}nonlocally causal\textquotedblright{}
(e.g.\ as an instantaneous \textquotedblleft{}common cause\textquotedblright{}),
or are characterized only as \textquotedblleft{}a-causal\textquotedblright{}
or \textquotedblleft{}formal\textquotedblleft{}.

\subsection{Instantaneous influences associated with nonlocal correlations are,
at a minimum, near-instantaneous, superluminal influences}

The instantaneous influences usually associated with observation in
EPR-type experiments of nonlocal correlations would be influences,
again whether causal or a-causal (see 1.1.), that operate, by definition,
at an infinite velocity. However, as is well-known, the assumption
of infinite velocity could never be proven experimentally. In other
words, the question appears to be undecidable by any conceivable experiment
of whether or not an observation, such as the appearance of nonlocal
correlations, is in fact due to an infinite-velocity influence (i.e.\ an
\textquotedblleft{}instantaneous correlation\textquotedblright{}).
By contrast, the generation of positive evidence for finite-velocity,
yet superluminal, influences is possible by way of scientific experimentation,
at least in principle (e.g.\ see Salart \textit{et al.}~\cite{Salart.2008spacelike};
Cocciaro \textit{et al.}~\cite{Cocciaro.2011lower}). In any case,
what can, and indeed has been, established by way of EPR-type experiments
is that influences exist in nature which, if they are not instantaneous,
then they at least must be superluminal (however, they must not be
signalling, of course). Consequently, we here take a conservative
position and assume that nonlocal correlations could either be instantaneous
correlations or near-instantaneous, superluminal correlations; again,
the reason is that proof of instantaneity is beyond scientific measurement.

\subsection{Nonlocal correlations are unpredictable but not necessarily intrinsically
random correlations}

It is well-known to physicists that no mathematical proof exists which
could decide \textquotedblleft{}whether or not a given series of digits
is in fact random or only seems random\textquotedblright{} (Chaitin~\cite{Chaitin.1987information}).
This does not mean, of course, that powerful statistical tests are
unavailable for demonstrating the lack of \textbf{certain} patterns
in a digital sequence; nevertheless \textquotedblleft{}no finite set
of tests can ever be considered complete, as there may be patterns
not covered by such tests\textquotedblright{} (Pironio \textit{et
al.}~\cite{Pironio.2010random}). In short, it is undecidable \textendash{}
as a matter of principle again \textendash{} whether a digital sequence,
for example, a data sequence obtained from an EPR-type experiment,
is genuinely random or merely pseudorandom. The principal undecidability
of whether data sequences derived from quantum experiments are either
random or merely pseudorandom has the following consequence: no experiment
could refute the possibility that deterministic processes might rule
at all levels of physical reality, including at the level of the quantum.
Therefore, while indeterminism remains a metaphysical assumption which
is greatly preferred over determinism by many quantum physicists,
it must be noted also that indeterminism is neither a conclusive experimental
fact nor a logical requirement on the basis of existing evidence.
Importantly, in relation to EPR-type experimental findings, the two
competing metaphysical assumptions \textendash{} determinism and indeterminism
\textendash{} share an all-important feature: the \textbf{statistical
unpredictability} of measurement outcomes. Therefore, given the above-described
undecidability concerning the existence, or not, of free randomness,
we here take again a conservative position, and we characterize EPR-type
correlation phenomena as merely \textquotedblleft{}unpredictable\textquotedblright{}
instead of as a sure sign of \textquotedblleft{}intrinsic randomness\textquotedblright{}. 

For example, statistical unpredictability may be the result of intrinsic
complexity rather than intrinsic randomness. That is, unpredictability
could be entirely a function of deterministic processes, i.e.\ at
the level of individual events predictability would be impossible
due to lack of precise knowledge. Famously, in deterministic chaos,
(long-term) unpredictability of behavior is due to \textbf{intrinsic
complexity}, and not due to intrinsic randomness. Therefore, in the
case of complexly-structured, deterministic systems an assumption
of free randomness is not required to account for the appearance of
statistical unpredictability of measurement outcomes. 

Therefore, it should remain an open question in quantum physics whether,
or not, the correlations that are observed in EPR-type experiments
do in fact represent \textbf{intrinsically random}, \textbf{instantaneous}
correlations. From the alternative perspective of deterministic theory,
we here argue that the quantum phenomenology as established by EPR-type
experiments merely indicates that these correlations are \textbf{unpredictable},
not intrinsically random, and that they either may be \textbf{near-instantaneous},
\textbf{superluminal} or \textbf{instantaneous} (1.1. to 1.3.).

\section{Evaluating the validity of the non-signalling theorem in generalizations
of Bell\textquoteright{}s theorem}

During the last decade, as was already mentioned in the Introduction,
an argument has gained traction, especially in work concerning the
foundations of quantum cryptography, which uses the validity of the
non-signalling theorem, to generalize the predictions of Bell\textquoteright{}s
theorem i.e.\ to rule out the possibility of NHVTs including de Broglie-Bohm
theory. In summary, the argument has three stages: (1) assume the
validity of the non-signalling theorem as a foundational principle
concerning the nature of physical reality, (2) if so, then the unpredictability
of nonlocal correlations as observed in EPR-type experiments is essentially
due to intrinsic randomness, and therefore (3) deterministic approaches
to quantum mechanics, including even NHVTs, are impossible. A critique
of this argument will be the main topic for the remainder of this
article.

\subsection{On the use of the non-signalling theorem in the foundations of quantum
cryptography}

There is a wide-spread belief, or basic assumption, especially in
the growing literature on quantum cryptography, that the non-signalling
theorem guarantees the presence of \textquotedblleft{}free randomness\textquotedblright{}
i.e.\ of fundamental indeterminism. Conversely, the belief is wide-spread
also that a deterministic approach, in combination with nonlocality,
necessarily must imply superluminal signalling. For example, one representative
statement from that literature reads \textquotedblleft{}any state
that is deterministic and nonlocal allows signalling\textquotedblright{}
(e.g.\ Barrett \textit{et al.}~\cite{Barrett.2005no}). However,
if that statement were true in an ultimate sense, then Bell\textquoteright{}s
theorem would not only negate the possibility of LHVTs but also of
any NHVTs such as de Broglie-Bohm theory. Traditionally, however,
compatibility of NHVTs and the predictions of Bell\textquoteright{}s
theorem had always been assumed (e.g.\ Bell~\cite{Bell.1987speakable};
Bohm and Hiley~\cite{Bohm.1993undivided}; Holland~\cite{Holland.1993};
Valentini~\cite{Valentini.2002signal-locality}). Consequently, there
now appears to have surfaced a fundamental contradiction between the
well-known predictions of Bell\textquoteright{}s theorem, on the one
hand, and the recent claims in the literature on quantum cryptography
concerning the use of the non-signalling theorem as a foundational,
instead of as an operational, principle. As evidence for the \textquotedblleft{}new\textquotedblright{}
incompatibility we will next discuss a number of statements by the
authors who were mentioned in the Introduction already (Gallego \textit{et
al.}~\cite{Gallego.2013full}). We will point out, in appropriate
places, that the presumption of incompatibility results from the assumption
that the non-signalling principle serves as an ontic, foundational
principle, which, as we will explain, presents a view of the principle
that has lost its essential connection to epistemological concerns. 

According to Gallego \textit{et al.}~\cite{Gallego.2013full}, the
non-signalling principle \textquotedblleft{}states that no instantaneous
communication is possible, which in turn imposes a local structure
on events, as in Einstein\textquoteright{}s special relativity.\textquotedblright{}
In other words, any kind of superluminal signalling or communication
is prohibited, whether by instantaneous or quasi-instantaneous, superluminal
influences. (For the remainder of this analysis we will only discuss
possible consequences of instantaneous influences in the context of
non-signalling.) We fully agree with Gallego \textit{et al.}~\cite{Gallego.2013full}
that this indeed is the essential meaning of the non-signalling principle.
Put differently, the impossibility by epistemic agents to signal each
other by way of quantum nonlocality protects against violations of
special relativity. However, we question the basic assumption by Gallego
\textit{et al.}~\cite{Gallego.2013full} that the non-signalling
principle applies also to instantaneous influences that are beyond
control for the purposes of communication or signalling. Crucially,
the non-signalling principle does not state \textquotedblleft{}no
instantaneous influences are possible\textquotedblright{}, but it
only states \textquotedblleft{}no instantaneous communication is possible\textquotedblright{};
otherwise, for example, the formation of nonlocal correlations in
EPR-type experiments would be impossible to account for, or even to
talk about, in terms of possible explanations in physics (compare
1.1.). Importantly, we point out that as long as control is impossible
of instantaneous influences, these influences, even if acting instantaneously,
do not violate the non-signalling constraint. In other words, if such
influences are informationally inaccessible to epistemic agents, i.e.\ if
influences are genuinely \textquotedblleft{}hidden influences\textquotedblright{}
(e.g.\ \textquotedblleft{}hidden variables\textquotedblright{}),
then instantaneous influences by way of nonlocal hidden variables
do not have to contradict special relativity. This is essentially
what is implied by an epistemic, operational, or phenomenological
interpretation of the non-signalling theorem. 

Valentini~\cite{Valentini.2002signal-locality}, for example, showed
that \textquotedblleft{}all deterministic hidden-variable theories,
that reproduce quantum theory for a \textquoteleft{}quantum equilibrium\textquoteright{}
distribution of hidden variables, predict the existence of instantaneous
signals at the statistical level for hypothetical \textquoteleft{}nonequilibrium
ensembles\textquoteright{}.\textquotedblright{} Crucially, however,
the assumption of \textquotedblleft{}instantaneous signals at the
statistical level\textquotedblright{} for quantum theories sharing
both determinism and nonlocality does not mean that such theories
would have to contradict the non-signalling theorem. For example,
after considering the possibility, between two distant members of
a correlated pair of spin-\textonehalf{} particles, \textquotedblleft{}of
nonlocal information flow at the hidden-variable level\textquotedblright{},
Valentini~\cite{Valentini.2002signal-locality} explained: \textquotedblleft{}Of
course, in equilibrium this information flow is not visible at the
statistical level, because as many outcomes flip from +1 to -1 as
from -1 to +1.\textquotedblright{} An epistemic, phenomenological
interpretation of the non-signalling theorem is therefore implicit
in the argument by Valentini~\cite{Valentini.2002signal-locality}:
\textquotedblleft{}It is as if there is a conspiracy in the laws of
physics that prevents us from using nonlocality for signalling. But
another way of looking at the matter is to suppose that our universe
is in a state of statistical equilibrium at the hidden-variable level,
a special state in which nonlocality happens to be hidden. The physics
we see is not fundamental; it is merely a phenomenological description
of an equilibrium state.\textquotedblright{}

However, it is apparent that Gallego \textit{et al.}~\cite{Gallego.2013full}
have excluded, as of course have many others (e.g.\ Barrett \textit{et
al.}~\cite{Barrett.2005no}; Kofler \textit{et al.}~\cite{Kofler.2006experimenters};
Pironio \textit{et al.}~\cite{Pironio.2010random}; Colbeck and Renner~\cite{Colbeck.2011no,Colbeck.2012free}),
as a possibility an epistemic, phenomenological interpretation of
the non-signalling constraint: \textquotedblleft{}In fact, Bohm\textquoteright{}s
theory is both deterministic and able to produce all quantum predictions,
but it is \textbf{incompatible with no-signalling at the level of
hidden variables}.\textquotedblright{} And, they continue: \textquotedblleft{}Thus,
we assume throughout the validity of the no-signalling principle.\textquotedblright{}
From this last statement it becomes clear that these authors, like
many before them, have committed to a view of the non-signalling theorem
which has eliminated any vital reference to epistemic agents, and
to what epistemic agents can or can\textquoteright{}t do; they have
\textquotedblleft{}ontologized\textquotedblright{}, so to speak, a
principle derived solely on the basis of epistemic considerations
i.e.\ signalling or non-signalling. In consequence, the \textquotedblleft{}non-signalling\textquotedblright{}
principle was tacitly turned into a \textquotedblleft{}no hidden-influence\textquotedblright{}
principle. The justification for this move and for transforming the
very meaning of non-signalling has not yet been revealed. Is this
a valid interpretation of the non-signalling principle? Can the notion
of \textquotedblleft{}non-signalling\textquotedblright{} really be
reduced to a notion of \textquotedblleft{}non-influencing\textquotedblright{}? 

Regarding the problem of \textquotedblleft{}non-signalling\textquotedblright{}
versus \textquotedblleft{}non-influencing\textquotedblright{}, specifically
in relation to NHVTs such as de Broglie-Bohm theory, for example,
Holland~\cite{Holland.1993} commented: \textquotedblleft{}To summarize,
the quantum potential implies that a certain kind of \textquoteleft{}signalling\textquoteright{}
does, in fact, take place between the sites of distantly separated\ldots{}
particles in an entangled state, if one of the particles undergoes
a local interaction. This transfer of information cannot, however,
be extracted by any experiment which obeys the laws of quantum mechanics.\textquotedblright{}
This is in agreement with the above-cited position by Valentini~\cite{Valentini.2002signal-locality}
who explained that in the manifestation of nonlocal correlations \textquotedblleft{}information
flow is not visible at the statistical level\textquotedblright{};
therefore control of information flows by epistemic agents for the
purposes of signalling is again prohibited. Finally, Holland~\cite{Holland.1993}
concluded that therefore \textquotedblleft{}\ldots{}the failure of
quantum correlations to provide a signalling mechanism at the empirical
level is consistent with the requirement of special relativity that
no signal be transmitted faster than the speed of light.\textquotedblright{}

\subsection{Circular reasoning in the relationship between the non-signalling
principle and the assumption of intrinsic randomness}

\begin{figure}
\begin{centering}
\includegraphics[width=0.75\columnwidth]{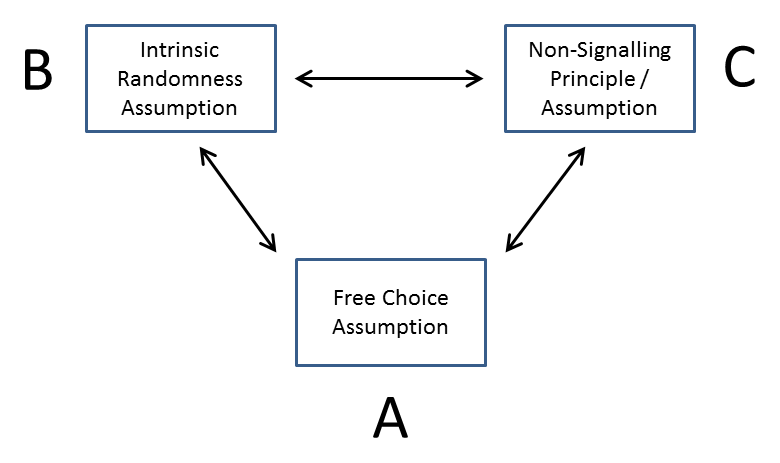}
\par\end{centering}

\caption{Relational diagram showing the interdependency of three basic assumptions
(A--C) which are discussed, for example, in Colbeck and Renner~\cite{Colbeck.2011no,Colbeck.2012free},
and Gallego \textit{et al.}~\cite{Gallego.2013full}. The validity
of each of these assumptions is central to the conclusion that deterministic
theories, including those based upon nonlocal hidden variables (i.e.\ de
Broglie-Bohm theory) are impossible because they contradict the non-signalling
principle.\label{fig:1}}
\end{figure}

We will next describe another argument against the use of the non-signalling
principle as an ontic foundational principle in favor of the existence
of intrinsic randomness, in addition to the argument already presented
above (2.1.). This second argument is based on the following simple
observations: it is exactly because standard quantum theory assumes
the unpredictability of nonlocal correlations to be evidence of intrinsic
randomness that these correlations are presumed to be \textquotedblleft{}intrinsically
non-signalling\textquotedblright{} in the first place; it is equally
valid therefore to say that the non-signalling assumption (C in Fig.~\ref{fig:1})
is a consequence of the prior free (intrinsic) randomness assumption
(B in Fig.~\ref{fig:1}). 

Similarly to the first step of the argument by Gallego \textit{et
al.}~\cite{Gallego.2013full} which did correctly recognize the inseparability,
or relationality, of assumptions A and B (see bidirectional arrow
between B and A in Fig.~\ref{fig:1}), the second step of their argument
also, \textbf{as is pointed out in the present work only}, is equally
characterized by an inseparability, or relationality, of assumptions
B and C (see bidirectional arrow between C and B in Fig.~\ref{fig:1}).
Therefore, it is questionable whether the non-signalling principle
can be used to argue for the necessity of free randomness in nature,
as only the \textbf{prior} assumption of intrinsic randomness could
justify any foundational, physical implications of the non-signalling
theorem to begin with. In the final section of this article (2.3),
we will first provide a short (historical) overview of reasons for
introducing the non-signalling principle, and then briefly return
to discussion of the incomplete metaphysical status of the non-signalling
principle.

\subsection{The non-signalling theorem}

\begin{figure}
\begin{centering}
\includegraphics[width=0.75\columnwidth]{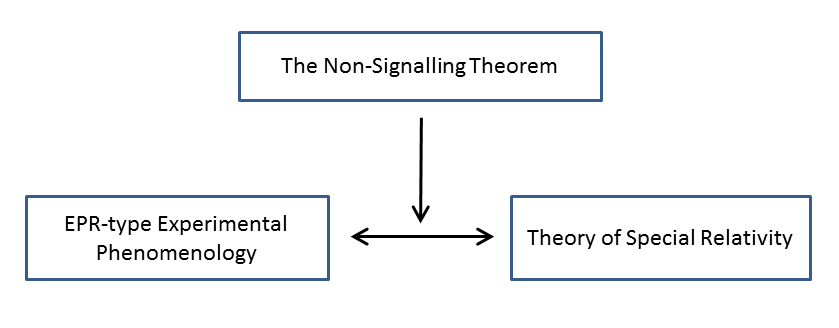}
\par\end{centering}

\caption{The non-signalling theorem or principle.\label{fig:2}}
\end{figure}

The non-signalling theorem was introduced, of course, to acknowledge
that EPR-type experiments, which indicate \textquotedblleft{}instantaneous
correlations\textquotedblright{} between space-like separated quantum
detectors, do not violate special relativity (e.g.\ Eberhard~\cite{Eberhard.1978bells}).
Special relativity imposes a fundamental constraint on all physical
phenomena whether quantum or classical. In particular, special relativity
imposes a fundamental limit on the maximum speed of any physical influences
existing in the vacuum of spacetime \textendash{} the speed of light.
Therefore, special relativity should outright prohibit the appearance
in nature of any kind of \textquotedblleft{}superluminal influences\textquotedblright{}
whether infinite-velocity influences (i.e.\ instantaneous influences)
or finite-velocity influences (e.g.\ near-instantaneous, superluminal
influences). However, as was described before, EPR-type experiments
have provided ample evidence for the existence of instantaneous, or
at least, superluminal influences (see 1.3). Such influences are apparent
from the measurement of physical effects which manifest -- at velocities
far exceeding the speed of light -- between distant quantum detectors.
Again, how could the existence of faster-than-light influences be
compatible with special relativity? In regards to EPR-type experimental
phenomenology, the tension which arises as a function of the \textbf{apparent}
incompatibility between special relativity, on the one hand, and the
observation of instantaneous, or at least, superluminal correlations,
on the other hand, is relieved through the introduction of the non-signalling
theorem or principle (see Fig.~\ref{fig:2}).

Clearly, the need to invoke non-signalling as a theorem for avoiding
conflict with special relativity reveals just how extraordinary the
phenomenology is of EPR-type experiments. This introduction is all
the more remarkable as the theory of special relativity puts firm
constraints already on the nature of allowable micro-causal processes,
and yet the non-signalling principle itself does not refer to causal
processes at all. The non-signalling principle refers to signals only,
or rather their absence, i.e.\ to epistemic \textquotedblleft{}states
of knowledge\textquotedblright{} rather than to any ontic physical
or causal states. The fact that an \textbf{epistemological} rather
than an ontological argument suffices to resolve the apparent incompatibility
between special relativity and EPR-type data has astonished many thinkers
on quantum foundations. This includes for example John Bell also who
asked (Bell~\cite{Bell.2004nouvelle}): 

\textquotedblleft{}Do we then have to fall back on \textquoteleft{}no
signalling faster than light\textquoteright{} as the expression of
the fundamental causal structure of contemporary theoretical physics?
This is hard for me to accept. For one thing we have lost the idea
that correlations can be explained, or at least this idea awaits reformulation.
More importantly, the \textquoteleft{}no-signaling\ldots{}\textquoteright{}
notion rests on concepts which are desperately vague, or vaguely applicable.
The assertion that \textquoteleft{}we cannot signal faster than light\textquoteright{}
immediately provokes the question: Who do \textbf{we} think we are?
\textbf{We} who can make \textquoteleft{}measurements\textquoteright{},
\textbf{we} who can manipulate \textquoteleft{}external fields\textquoteright{},
\textbf{we} who can signal at all, even if not faster than light?
Do we include chemists, or only physicists, plants, or only animals,
pocket calculators, or only mainframe computers?\textquotedblright{} 

\begin{table}
\begin{centering}
\renewcommand*\arraystretch{2}%
\begin{tabular*}{1\columnwidth}{@{\extracolsep{\fill}}|>{\raggedleft}m{0.25\columnwidth}|>{\centering}m{0.33\columnwidth}|>{\centering}m{0.33\columnwidth}|}
\hline 
 & \textbf{Signalling / Non-signalling Influences} & \textbf{Instantaneous Influences}\tabularnewline[2mm]
\hline 
$\vphantom{\Biggl(\Biggr)}$Basic relations: & $\vphantom{\Biggl(\Biggr)}$Sender and Receiver & \textquotedblleft{}Unknown relations\textquotedblright{} \textsuperscript{1}\tabularnewline[2mm]
\hline 
$\vphantom{\Biggl(\Biggr)}$Operational ability: & Ability, or inability, by epistemic agents to control information
transfers & Registering random events \tabularnewline[2mm]
\hline 
$\vphantom{\Biggl(\Biggr)}$Influence is: & Purposeful;

a \textquotedblleft{}message\textquotedblright{} is, or is not, conveyed & Not purposeful; blind\tabularnewline[2mm]
\hline 
$\vphantom{\Biggl(\Biggr)}$Available must be: & Epistemic agents \textbf{and} physical processes & Physical processes\tabularnewline[2mm]
\hline 
$\vphantom{\Biggl(\Biggr)}$Conceptual framework: & $\vphantom{\Biggl(\Biggr)}$Epistemology \textbf{and} Ontology & Ontology\tabularnewline[2mm]
\hline 
\end{tabular*}
\par\end{centering}

\caption{Distinguishing signalling/non-signalling influences from instantaneous
influences. \textsuperscript{1}The reference in the Table to \textquotedblleft{}unknown
relations\textquotedblright{} only states the obvious namely the lack
of scientific understanding of the \textbf{physical nature} of \textquotedblleft{}instantaneous
influences\textquotedblright{}. For example, an instantaneous influence
does neither imply \textquotedblleft{}efficient causation\textquotedblright{}
nor any kind of \textquotedblleft{}action-reaction\textquotedblright{}-type
scenarios unless when considering a wholly \textbf{operational} meaning
of an \textquotedblleft{}action-reaction\textquotedblright{} relation.
For example, in the context of instantaneous influences, as part of
an EPR-type experiment, an \textquotedblleft{}action\textquotedblright{}
may be the defining of a measurement setting in location A, whereas
a \textquotedblleft{}reaction\textquotedblright{} might refer to the
physical measurement effect at the quantum detector in location B;
again operationally only.\label{tab:1}}
\end{table}
We conclude that the metaphysical status of the non-signalling principle
still remains a very much open question even today. For example, as
was alluded to above many times already, it is still undecided whether
the non-signalling principle is a foundational principle, or an operational,
phenomenological one. We conclude that until this all-important question
is answered, the argument, for example, by Gallego \textit{et al.}~\cite{Gallego.2013full}
must be considered, at a minimum, to be incomplete concerning their
claim that the non-signalling principle is a conclusive argument in
favor of intrinsic randomness in nature (compare Fig.~\ref{fig:1},
B and C). 

Importantly, our analysis maintains that while the non-signalling
principle prohibits faster-than-light communication it does not prohibit
the possibility of instantaneous influences. Put differently, in combination
with the non-signalling principle, the meaning of special relativity
is changed from an absolute \textquotedblright{}no influence can exist
which is faster than the speed of light\textquotedblright{} to the
less restrictive statement that \textquotedblright{}no instantaneous/
superluminal influence can exist \textbf{that can be controlled for
signalling purposes}\textquotedblright{}. Critically, the conflation
in meaning of the notions of \textquotedblleft{}signalling / non-signalling
influence\textquotedblright{} and \textquotedblleft{}instantaneous
influence\textquotedblright{} has led to much confusion in the literature,
and worse, to misleading conclusions. The Table summarizes differences
for better overview. 

The present work sought to argue against proposals that seek to construe
the non-signalling theorem as a \textquotedblleft{}non-influence theorem\textquotedblright{}.
Specifically, we have argued that the non-signalling theorem serves
to eliminate only the capacity of epistemic agents to signal information
by way of instantaneous, or near-instantaneous, influences. However,
the non-signalling theorem does not equal a non-influence theorem,
e.g.\ the non-signalling theorem does not prohibit the existence
of instantaneous influences, for example, in the formation of nonlocal
correlations as part of EPR-type experiments (see 1.1. to 1.3.). 

Finally, we conclude that \textquotedblleft{}epistemic non-signalling\textquotedblright{}
is likely to be sufficient to ensure the peaceful coexistence of special
relativity and quantum nonlocality. A major reason in favor of our
conclusion is that any definition of signalling or non-signalling
invariably requires a reference to epistemic agents, and what these
agents can actually measure and report (see Table~\ref{tab:1}). 
\begin{acknowledgments}
Work by Jan Walleczek at Phenoscience Laboratories (Berlin) is partially
funded by the Fetzer Franklin Fund of the John E. Fetzer Memorial
Trust (\url{http://www.fetzer-franklin-fund.org/}). Work by Gerhard
Grössing at the Austrian Institute for Nonlinear Studies is also partially
funded by the Fetzer Franklin Fund of the John E.\ Fetzer Memorial
Trust.
\end{acknowledgments}

\providecommand{\href}[2]{#2}\begingroup\raggedright\endgroup

\end{document}